# Opto-Acoustic Biosensing with Optomechanofluidic Resonators

Kaiyuan Zhu[1], Kewen Han[1], Tal Carmon[2], Xudong Fan[3] and Gaurav Bahl[1†]

[1]Department of Mechanical Science and Engineering, The University of Illinois at Urbana-Champaign, Urbana, Illinois 61801 USA
[2]Department of Biomedical Engineering, University of Michigan, Ann Arbor, MI 48109, USA
[3]Mechanical Engineering, Technion - Israel Institute of Technology, Israel
[†]bahl@illinois.edu

**Abstract**

Opto-mechano-fluidic resonators (OMFRs) are a unique optofluidics platform that can measure the acoustic properties of fluids and bioanalytes in a fully-contained microfluidic system. By confining light in ultra-high-Q whispering gallery modes of OMFRs, optical forces such as radiation pressure and electrostriction can be used to actuate and sense structural mechanical vibrations spanning MHz to GHz frequencies. These vibrations are hybrid fluid-shell modes that entrain any bioanalyte present inside. As a result, bioanalytes can now reflect their acoustic properties on the optomechanical vibrational spectrum of the device, in addition to optical property measurements with existing optofluidics techniques. In this work, we investigate acoustic sensing capabilities of OMFRs using computational eigenfrequency analysis. We analyze the OMFR eigenfrequency sensitivity to bulk fluid-phase materials as well as nanoparticles, and propose methods to extract multiple acoustic parameters from multiple vibrational modes. The new informational degrees-of-freedom provided by such opto-acoustic measurements could lead to surprising new sensor applications in the near future.

## 1. Introduction

Optofluidic technologies conveniently integrate microfluidics and photonics on the same device and therefore provide a unique capability to manipulate and detect optically responsive analytes [1-4]. In such platforms, various optical parameters can be used as the sensing signal, such as refractive index [5-7], fluorescence (i.e. spontaneous emission), laser emission (i.e. stimulated emission) [8], and Raman scattering [2, 9]. Many different optofluidic structures have been explored, such as surface plasmon resonance devices [10], ring resonators [5, 11-13], liquid core capillaries [14], microtoroids [15], bubbles [16-18], droplets [19, 20], photonic crystals [21, 22], liquid core waveguides [2, 3, 7, 23], etc. which enable sensitive detection of DNAs, proteins, viruses, cells, and bacteria down to the level of single molecules or particles [15, 21, 24]. In addition, optical nanoparticle sizing and sorting [25-28] as well as flow control [29, 30] have also been achieved.

Similar to the manner in which we employ audio cues received through our ears to obtain information about our surroundings and to detect invisible hazards, the use of sound waves and vibrations for biosensing can provide details that are not accessible by other methods. For instance, the measurement of the frequency shift of a mechanical resonator to infer mass of a particle has been achieved with both fluid immersed [31] resonators and with fluid infused suspended microchannel resonators (SMR) [32]. The SMR method has recently achieved mass detection resolution as impressive as 0.85 attograms [33]. SMRs have also been employed for a wide range of biological applications such as monitoring cell growth [34] and studying single cell biophysical properties [35]. Expanding beyond this concept, the transmission and reflection of acoustic waves shorter than the dimensions of a cell can serve in revealing details on the membrane stiffness and cytoskeleton that cannot be observed by conventional microscopy. Indeed, high-frequency opto-acoustic interactions have already allowed us to measure fluid properties [36, 37], hydration of biofilms [38], and image mechanical properties of bioparticles beyond the abilities of optical microscopy [39, 40].

The emergence of optomechanics techniques that link mechanical vibrations to optical fields and forces (radiation pressure, gradient force, electrostriction) [41-43] has provided a tool for optically actuating and measuring vibrations from kHz to GHz frequencies. In the case of radiation pressure oscillation, for instance, laser light exerts centrifugal pressure on the walls of a resonator, thereby deforming it slightly, which modifies the optical modes of the resonator. This in turn affects the laser power coupling into the resonator and creates feedback. When the gain of this feedback loop is sufficient, this can result in the parametric actuation of mechanical vibrations like breathing modes and wineglass modes as shown in previous experiments [44, 45]. Mechanical oscillation of the device modulates the optical modes and the optical path length, thereby generating sidebands of the input optical signal. The vibrational frequency and mechanical power spectrum are then determinable through heterodyned measurement of these optical signals with a photodetector. Recently, the frequency regime for such mechanical interaction has been extended into the 10's of GHz using stimulated Brillouin scattering [46-48].

Our recent development of opto-mechano-fluidic resonators (OMFRs) [44, 45], inspired simultaneously by optomechanics, optofluidics, and SMRs, has given optomechanical systems the ability to interact with arbitrary fluids. OMFRs are silica microfluidic capillaries of about 100 μm diameter and several cm length, having a central 'bottle' region where optical and mechanical modes can be simultaneously confined while a fluid flows through. Similar to SMRs, the confinement of liquid analytes within OMFRs prevents acoustic energy from leaving the resonator and enables very high mechanical mode quality factors. In contrast to SMRs, the mechanical modes in OMFRs are able to entrain the liquids contained inside, forming a hybrid eigenmode where the fluid and shell are coupled. Such hybrid modes can be harnessed either through radiation pressure or through stimulated Brillouin scattering as we have demonstrated recently [44]. Further, by optomechanically measuring the mechanical resonance frequency, it is possible to measure the density and speed of sound [44, 45], and viscosity [49] of the entrained fluid. In this paper we computationally study the sensitivity of OMFR vibrational modes for optically measuring such physical properties of fluids and biologically relevant micro/nano-particles. We also discuss techniques to uniquely determine the various acoustic properties of the entrained fluids using measurements of multiple vibrational modes.

## 2.    Mass sensing capabilities of optomechanofluidic resonators

### 2.1    Numerical simulation setup

The OMFR consists of two physics domains, a solid-phase shell made of fused-silica glass and the fluid contained inside. The capillary shell is cylindrical with a bottle-like region that can be quantified by a radius of curvature β. The capillary is widest at the equator with an outer diameter $D$ and inner diameter d. An aspect ratio parameter, $\alpha = d/D$, determines the wall thickness of the capillary. In this study, we are most interested in "breathing" vibrational modes and wineglass modes that are confined at the equator. Thus, the length, L, of the simulated region is determined such that boundary conditions at the two ends of the capillary will not affect the mechanical eigenfrequencies of the equatorially confined vibrational modes. In our simulations, the OMFR has an outer diameter D = 100 μm, radius of curvature β = 5000 μm,

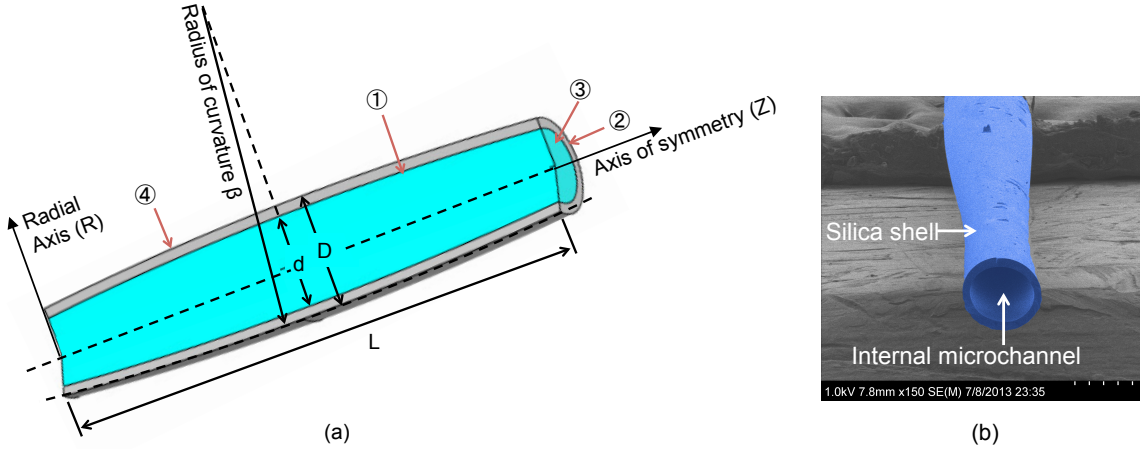

*Figure 1 (a) **Geometry and boundary conditions for the OMFR.** Geometry parameters are outer diameter of the shell D, inner diameter of the shell d, length L, aspect ratio α=d/D, radius of curvature β; Boundary conditions: ① Pressure loads to the shell and normal accelerations to fluid phase are applied at the fluid-solid interface, ② fixed boundary condition at the ends of the solid shell, ③ sound hard boundary condition at the ends of the fluid phase (normal acceleration vanishes) and ④ free boundary condition at outer surface of the shell. (b) Scanning electron micrograph of an OMFR showing the shell and microchannel.*

aspect ratio $\alpha$ = 0.85, and length L = 600 μm. Fig.1 shows the geometry as well as boundary conditions applied to the OMFR simulations in this paper.

Eigenfrequency analysis on the OMFR is performed through finite element analysis using COMSOL Multiphysics simulation software (http://www.comsol.com). The governing equation for eigenfrequency study of the solid-domain displacement field is

$$-\rho_s \omega^2 \mathbf{u} - \nabla \sigma = \mathbf{F}_V \quad (1)$$

where $\omega$ is the eignfrequency of certain mode, **u** is the displacement tensor of the wall, $\sigma$ is the stress, and $\mathbf{F}_V$ is the vectorial force per unit volume resulting from the pressure of the fluid within the capillary. Density $\rho_s$ and elastic modulus $E$ are the two input material properties for the solid mechanics eigenfrequency study. Free boundary conditions are applied everywhere on the outer surface of the shell. Boundary conditions of the far ends of the capillary are not important here since only locally confined modes near the optical modes are considered. In the simulation, fixed boundary conditions are applied to the two ends of the wall.

Eigenfrequency analysis of the fluid domain is based on the wave equation of the sound wave,

$$\nabla \cdot \left(-\frac{1}{\rho_f}(\nabla p)\right) - \frac{\omega^2 p}{\rho_f c^2} = Q \quad (2)$$

where Q represents the acoustic source term that releases energy into the system. It is indicated from Equation (2) that fluid density $\rho_f$ and speed of sound in the fluid $c$ are the two input material properties that affect the eigenfrequency response of the fluid domain. Yet again, boundary conditions to the fluid at the two ends of the capillary are not important because the length of the capillary is significantly larger than the size of the locally confined modes. In our simulations, "sound hard" boundaries, in which normal acceleration vanishes ($\mathbf{n} \cdot \mathbf{a_0} = 0$), are applied at the two ends. To enable interaction between the two physics domains, we apply specific boundary conditions at the shell-fluid interface. In the solid shell the fluid pressure ($p$) acts as a boundary load to the wall ($\boldsymbol{F_A} = -p \cdot \boldsymbol{n}$), where $\boldsymbol{n}$ is the outward-pointing unit normal vector seen from inside the fluid domain. This boundary load is related to the external source $\boldsymbol{F_V}$ in Equation (1) by $\boldsymbol{F_V} = \frac{F_A}{t}$, in which $t$ is the thickness of the shell. In the fluid domain, normal acceleration of the shell inner surface acts as a source term to Equation (2), resulting in $Q = -\boldsymbol{n} \cdot \boldsymbol{a_0}$, where $\boldsymbol{a_0}$ is the acceleration of the wall at the solid-fluid interface. Physically, this means that the silica shell radially releases energy to the fluid with an acceleration field $\boldsymbol{a_0}$.

## 2.2 Hybrid fluid-shell vibrational modes

Some examples of computationally evaluated shell-fluid vibrational modes are shown in Fig. 2.

| Mode Order (N,L,M) | Mode Shape (cross-section view) | | Mode Family |
|---|---|---|---|
| | Shell Displacement Field | Fluid Pressure Field | |
| (1,1,0) | | | Breathing Modes |
| (2,1,0) | | | Breathing Modes |
| (1,2,0) | | | Breathing Modes |
| (2,2,0) | | | Breathing Modes |
| (1,1,2) | | | Wineglass Modes |
| (1,1,3) | | | Wineglass Modes |
| (1,1,4) | | | Wineglass Modes |

*Figure 2 **Mode shapes of the OMFR vibrational modes.** N=radial order, L=axial order, M=azimuthal order. Breathing modes are shown in the cross-section views of the R-Z plane. Wineglass modes (M ≠ 0) are shown in the cross-section views at the equator (Z=0 plane). Colors represent relative levels of displacement field in solid domain (red is high, blue is low), and represent relative levels of acoustic pressure in fluid domain (red is positive, blue is negative). Black solid lines represent the undeformed boundaries of the capillary. Although breathing modes with different radial orders (N) but the same axial orders (L) share nearly identical displacement fields, their acoustic pressure fields are different, which results in distinct trends in the frequency vs. fluid density curves (see Fig.3).*

Mode orders holding non-negative integer values are introduced to systematically study modal characteristics; radial order (N), axial order (L), and azimuthal order (M) are used to describe complexity of the modes. In the azimuthal direction, M is interpreted as the number of antinodes. Breathing modes have M=0 since there is no oscillation along the azimuthal direction. All wineglass modes have N=1 since a pressure antinode only occurs at the solid-fluid interface. Comparing the breathing modes with different radial orders (N) but the same axial orders (L), one can find that two modes have nearly identical mode shapes in the displacement fields, but their acoustic pressure fields are different since the number of radial antinodes are not the same. This results in dissimilar trends in frequency vs. density curves (see Fig. 3).

## 2.3 Sensitivity of hybrid fluid-shell modes to fluid properties

We now investigate how fluid density and speed of sound in the fluid will affect the resonant frequencies of these hybrid modes by performing eigenfrequency studies in COMSOL Mutiphysics. Fig. 3 shows eigenfrequency vs. fluid density at constant speed of sound for breathing modes and wineglass modes. In this simulation, fluids inside the capillary are assumed to have 1500 m/s speed of sound, a value that is close to the speed of sound of water. These plots are obtained by varying the density from 400 kg/m$^3$ to 2200 kg/m$^3$. The results show that modes from different families have distinct frequency and sensitivity trend with respect to density. For instance, the curves of the N=1 breathing modes from show positive change of frequency with respect to increasing density while the curves of the N=2 breathing modes from show negative change in frequency. This disparity of eigenfrequency dependence on fluid density is due to different acoustic pressure distributions within the fluid (see Fig. 2). Wineglass modes (Fig. 3(c)) and N=1 breathing modes (Fig. 3(a)) share similar decreasing trends of eigenfrequency with respect to increasing density, but shapes of the curves from the two families are slightly different.

In a simplistic model, the vibrational eigenfrequency $f$ depends on the effective stiffness $k_{eff}$ and the effective mass $m_{eff}$ of the mode. Since the effective mass is proportional to the density of the vibrating elements, it results in a lowering of frequency with increasing density. It is also true, however, that the effective stiffness is related to the bulk modulus ($B = \rho_f c^2$) of the fluid, which indicates that the effective stiffness increases with increasing fluid density. Since these two effects oppose each other, the tuning of frequency with density is highly dependent on the shape of the vibrational mode, and can take on both increasing and decreasing trends as we see in Fig. 3 (a,b).

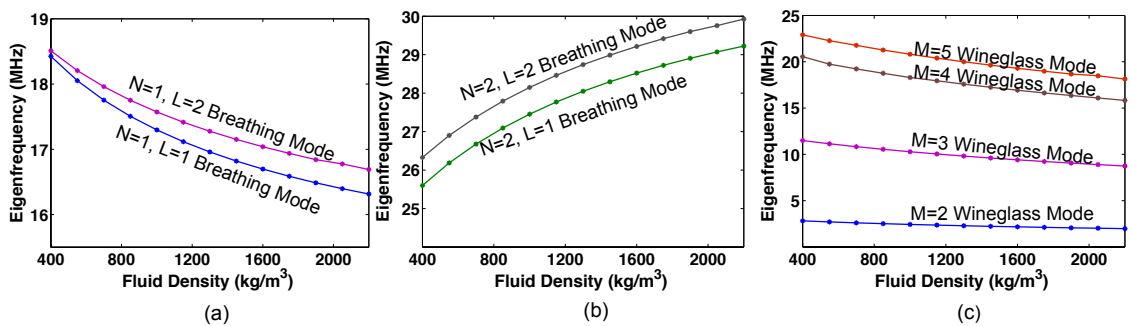

*Figure 3 **Variation of eigenfrequency with density** of the fluid assuming speed of sound of 1500 m/s for (a) first radial order (N=1) breathing modes (b) second radial order (N=2) breathing modes (c) wineglass modes with various azimuthal (M) orders. These plots show that hybrid modes from different families have distinct frequency and sensitivity trend with respect to density. Only the second radial order breathing modes show increasing frequency with increasing density.*

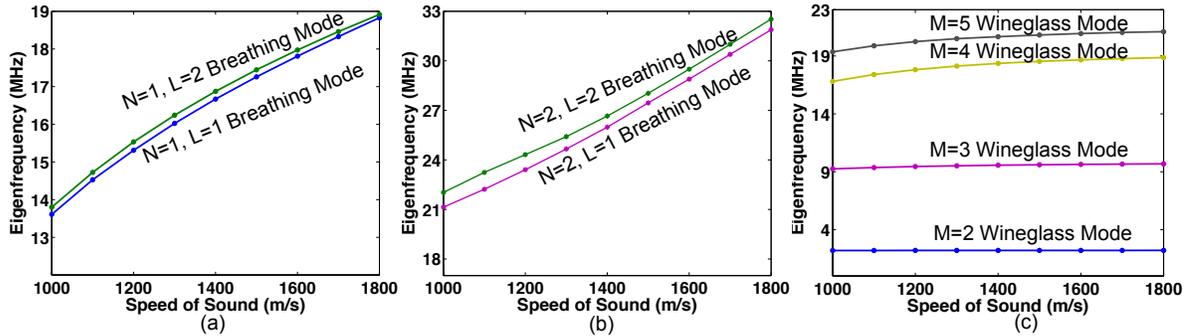

*Figure 4 **Variation of eigenfrequency with speed of sound** of the fluid assuming density of 1000 kg/m$^3$ for (a) first radial order (N=1) breathing modes (b) second radial order (N=2) breathing modes (c) wineglass modes with various azimuthal (M) numbers. These plots show that hybrid modes from different families exhibit similar increasing trends with respect to speed of sound.*

Plots of eigenfrequency vs. speed of sound at constant fluid density for various hybrid vibrational modes are shown in Fig. 4. These simulations are performed by assuming a constant fluid density of 1000 kg/m$^3$, a value close to the density of water, and by varying the speed of sound from 1000 m/s to 1800 m/s. These calculations show that modes from different families exhibit similar increasing trends with respect to speed of sound, which is consistent with the idea that bulk modulus increases at the same time.

In order to better understand how different fluid properties can affect the eigenmode and eigenfrequency of the OMFR, fluid density and speed of sound must simultaneously be varied to correspond to real fluids. As an example, Fig. 5(a,b) shows 3D surface plots of eigenfrequency as a function of fluid density and speed of sound for M=5, L=1 breathing and N=2, L=1 wineglass mode, respectively. It is easy to see that the two surfaces show very distinct shapes in this space and that at least one intersection contour exists. We can exploit such an intersection to uniquely determine the fluid properties based only on resonant frequency measurements of the two modes.

Let us suppose that one experimentally measures resonant frequencies of 20.8 MHz and 27.5 MHz from two distinct hybrid fluid-shell modes, and the experimenter knows that the lower frequency mode corresponds to an M=5 L=1 wineglass mode and the higher frequency mode is an N=2, L=1 breathing mode. Using the computed 3D surfaces for each mode (Fig. 5), we can draw a constant frequency contour at the measured resonant frequency. In other words, we are drawing a density vs. speed of sound plot at an eigenfrequency of 20.8 MHz on Fig. 5(a) and the same type of plot at an eigenfrequency of 27.5 MHz on Fig. 5(b). Next, the contours can be overlaid in the manner shown in Fig. 5(c). The unique intersection of the two contours shows a unique solution that simultaneously provides density and speed of sound of the fluid we are measuring inside the resonator. In this example, we obtain the fluid inside the capillary has density of 1000 kg/m$^3$ and sound speed of 1500 m/s.

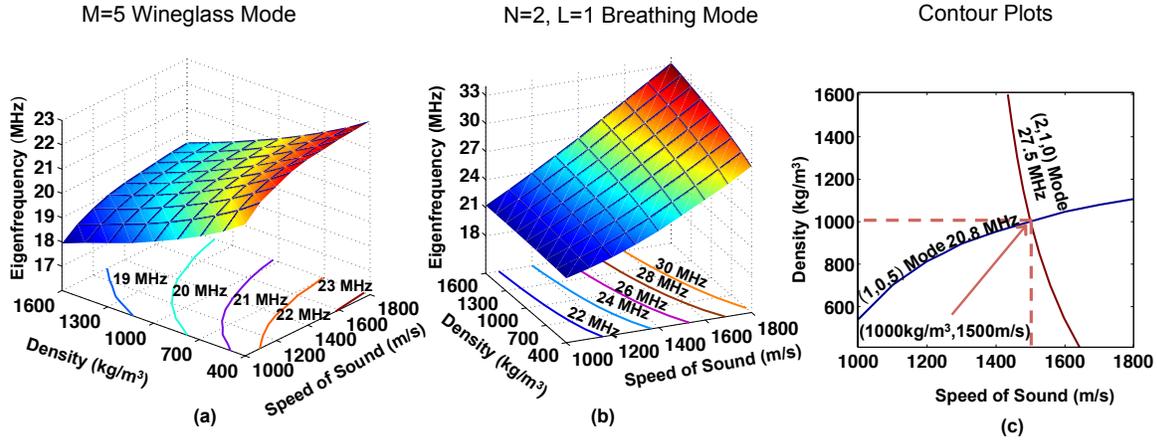

*Figure 5: (a) Frequency of the M=5 wineglass mode vs. fluid density and sound speed (b) Frequency of the N=2, L=1 breathing mode vs. fluid density and sound speed. For (a) and (b), contour plots with equal spacing frequencies are drawn at the bottom of each graph (c) Constant frequency contour plot of the N=2, L=1 breathing mode at 27.5 MHz and constant frequency contour plot of the M=5 wineglass mode at 20.8 MHz. The intersection is a unique solution that simultaneously provides density and speed of sound of the fluid we are measuring inside the resonator.*

### 2.4 Experimental and computational identification of hybrid fluid-shell vibrational modes and fluid properties

The correct identification of hybrid fluid-shell vibrational modes is essential for applying the method we describe in the previous section. Unfortunately, mode shapes in microdevices are challenging to directly observe, and an experimenter must rely on educated guesses. We provide here a simple method and experimental example to help identify the hybrid fluid-shell modes in OMFRs using a fluid of known density, and more accurately measure speed of sound of this fluid.

We take as a specific experimental example, an OMFR with outer diameter of 170 μm, an aspect ratio of 0.82 (15 μm shell thickness) and a radius of curvature of about 8000 μm. We infuse a NIST-calibrated standard oil into the OMFR having a density of 0.884 g/ml. Three resonant frequencies are experimentally measured at 11.54 MHz, 13.91 MHz, and 17.53 MHz (data presented in [49]). However, we do not yet have information on the speed of sound, nor do we know precisely which mode shapes these frequencies correspond to. For making this determination, we initially assume that the test fluid has speed of sound of 1300 m/s. Based on the known density and estimated speed of sound we can then calculate the possible vibrational modes whose eigenfrequencies are close to the experimentally measured frequencies. Next, for each shortlisted mode, we calculate the perturbation of eigenfrequency as a function of the speed of sound. As shown in Fig 6, where these calculated frequencies equal the measured frequencies we obtain likely candidates for both fluid speed of sound and the identified mode. However, since multiple solutions exist, we must rely on agreement between different measurements of the same fluid. Here (Fig. 6(a,b)) only two calculated modes corresponding to the two measured frequencies (11.5 MHz and 17.3 MHz) yield approximately the same value of the speed of sound. The imprecision in measuring speed of sound of the test fluid (1275 m/s to 1290 m/s) is due to inaccuracies in the geometry of the OMFR computational model since the dimensions are measured under a microscope. These errors can potentially be calibrated out. We conclude that the experimentally measured 11.54 MHz mode is an M=4, L=3 wineglass mode and the 17.53 MHz mode is an M=6, L=2 wineglass mode. As a secondary confirmation of our above conclusion, we find that the N=2, L=1 breathing mode is the only likely computational solution for the 13.91 MHz measurement. The point of intersection shown in Fig 6(c) provides a speed of sound of the fluid as 1240 m/s, which is close to our previous estimation and is also subject to OMFR geometry errors.

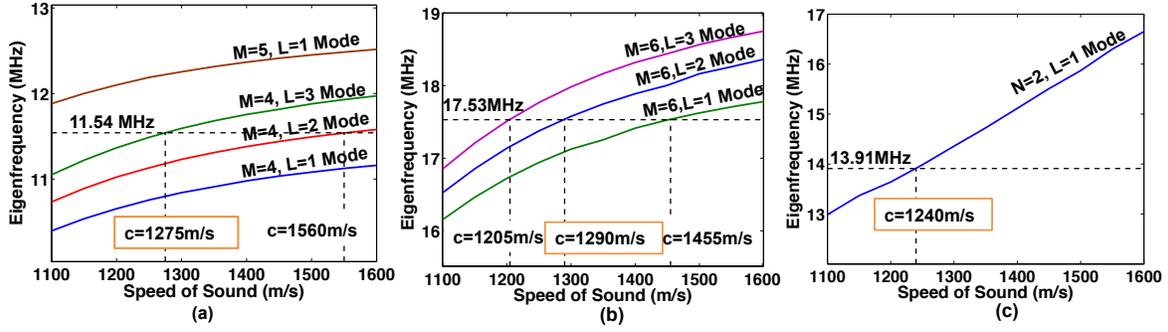

*Figure 6:* ***Computational identification of experimentally measured vibrational modes.*** *We compute the eigenfrequencies of candidate modes around our experimental measurements of (a) 11.54 MHz, (b) 17.53 MHz, and (c) 13.91 MHz using a test fluid with known density of 884 kg/m³ but unknown speed of sound. Here we conclude that the test fluid has speed of sound around 1275 m/s to 1290 m/s, and that the 11.54 MHz mode is an M=4, L=3 wineglass mode and the 17.53 MHz mode is an M=6, L=2 wineglass mode. The point of intersection in (c) from a N=2, L=1 breathing mode gives speed of sound of the fluid as 1240 m/s, which is consistent with the result obtained from (a) and (b) within a small margin of error.*

## 2.5     Sensitivity to micro/nano-particles

We now perform an analysis of the mass detection capability of OMFRs. We consider the presence of a fluid-suspended spherical particle within the OMFR. It is assumed that the particle is small enough such that it does not perturb the acoustic pressure distribution formed by the vibrational mode. The effective masses of vibrational modes of the OMFR are computationally evaluated, and perturbation theory is used to determine the frequency shift of the eigenfrequency upon replacing a small portion of the fluid with a small particle. We then define spatial mass sensitivity as the fractional shift of the eigenfrequency per mass of fluid replaced (in units of parts-per-million per nanogram) through the equation

$$\frac{1}{f}\frac{df}{dm} = \frac{1}{2}\left(\frac{|\bar{P}|}{max|P|}\right)^2 \quad (3)$$

where $\bar{P}$ is the local acoustic pressure at the particle location and $max|P|$ is the largest acoustic pressure inside the fluid. This mass sensitivity varies spatially over the resonator, and only depends on acoustic pressure at the location of the particle. To estimate the absolute frequency shift due to an analyte, one simply needs to multiply the mass sensitivity with the eigenfrequency of the mode and the total fluid mass replaced by the particle.

For illustration, we graphically plot in Fig. 7 the calculated sensitivity as a function of the location of the particle for N=1, L=1 breathing mode, N=1, L=2 breathing mode, and M=4 wineglass mode utilizing the geometric parameters defined in section 2.1. The plots show that greatest mass sensitivity for the breathing modes occurs at the center of the capillary (R=0), whereas the maximum mass sensitivity for the M=4 wineglass mode occurs only adjacent to the shell. However, the sensitivity in the wineglass case is 8 times greater than the maximum sensitivities of the breathing modes. As a specific example, the contrast of a silica microparticle ($\rho$=2203 kg/m$^3$) with a radius of 1 µm suspended in saline ($\rho$=1000 kg/m$^3$, $c$=1500 kg/m$^3$) at the center of the capillary is sufficient to perturb a 17.1 MHz (1,1,0) breathing mode of our simulated OMFR by about -7.6 Hz. This level of frequency shift can easily be detected by electronic spectrum analyzers.

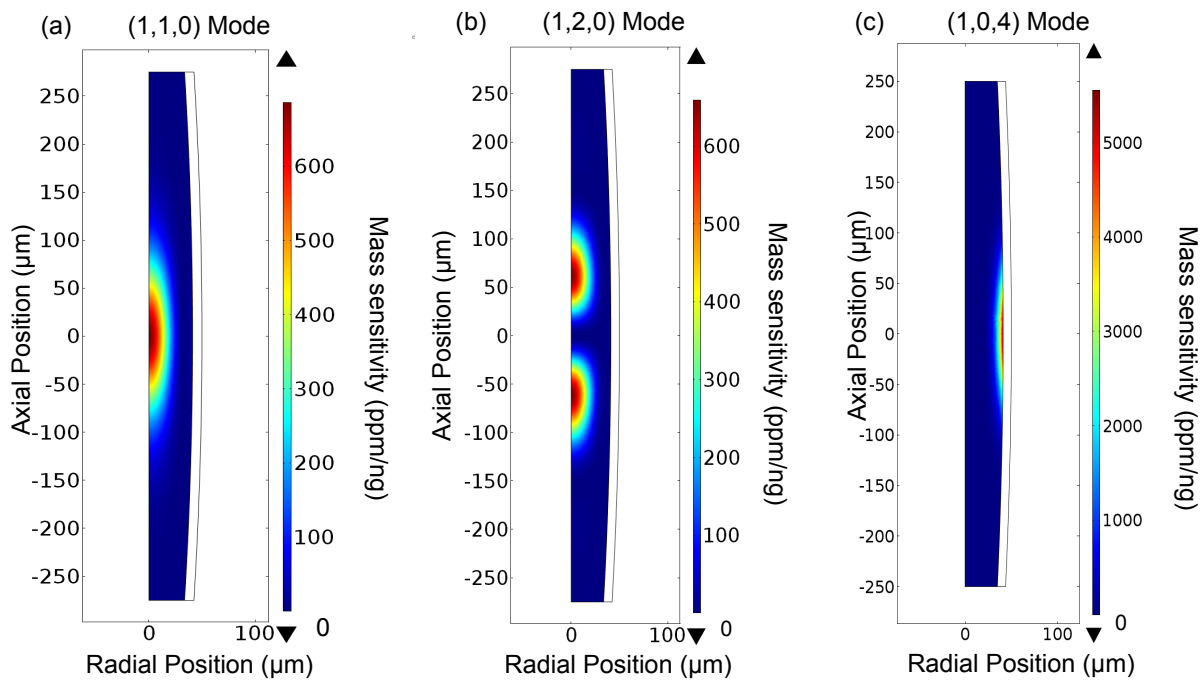

*Figure 7* **Graphical visualization of mass sensitivity of the OMFR hybrid fluid-shell vibrational modes** *to a particle that displace a volume of the fluid ($\rho$=1000 kg/m$^3$, $c$=1500 m/s) for (a) N=1, L=1 breathing mode, (b) N=1, L=2 Breathing Mode and (c) M=4 Wineglass Mode. Mass sensitivity is defined as fractional shift of the effective eigenfrequency per added mass (nanograms). The maximum sensitivity for the M=4 wineglass mode is 8 times greater than those of the two breathing modes, however, this sensitivity occurs only close to the shell.*

## 3.    Conclusion

In this work, we have investigated the mechanical sensing capabilities of the OMFR platform using multiphysics numerical eigenfrequency simulations. OMFR devices are shown capable of extracting the density and speed of sound of a few nanoliters of entrained fluid, using optical measurements of resonant vibrational frequencies of the hybrid fluid-shell modes. We apply this frequency measurement method to estimate the device sensitivity to single flowing cells or nanoparticles in the fluid. Finally, we demonstrate techniques by which experimental measurements of multiple modes can be employed to determine unique solutions for unknown properties of the contained fluids. Since the operational vibrational frequencies of a single OMFR can span the MHz - GHz regime [44, 45], OMFRs are a unique tool to achieve high-throughput analysis of the mechanical properties of fluids and bioanalytes over a wide range of timescales.

**Acknowledgments**

Funding for this research was provided through a University of Illinois at Urbana-Champaign Startup Grant and a UIUC Department of Mechanical Science and Engineering undergraduate research opportunities grant. We would also like to thank Prof. Randy Ewoldt, Prof. David Saintillan, Nathan Dostart, and Haden Duke for helpful discussions.